# ION DYNAMICS AT A RIPPLED QUASI-PARALLEL SHOCK: 2-D HYBRID SIMULATIONS

Yufei Hao[1,2], Quanming Lu[1,2], Xinliang Gao[1,2], and Shui Wang[1,2]

[1]CAS Key Laboratory of Geospace Environment, Department of Geophysics and Planetary Science, University of Science and Technology of China, Hefei 230026, China

[2]Collaborative Innovation Center of Astronautical Science and Technology, China

Corresponding author: Quanming Lu

Email:qmlu@ustc.edu.cn

# Abstract

In this paper, two-dimensional (2-D) hybrid simulations are performed to investigate ion dynamics at a rippled quasi-parallel shock. The results show that the ripples around the shock front are inherent structures of a quasi-parallel shock, and the reformation of the shock is not synchronous along the surface of the shock front. By following the trajectories of the upstream ions, we find that these ions behave differently when they interact with the shock front at different positions along the shock surface. The upstream particles are easier to transmit through the upper part of a ripple, and the bulk velocity in the corresponding downstream is larger, where a high-speed jet is formed. In the lower part of the ripple, the upstream particles tend to be reflected by the shock. For the reflected ions by the shock, they may suffer multiple stage acceleration when moving along the shock surface, or trapped between the upstream waves and the shock front. At last, these ions may escape to the further upstream or enter the downstream, therefore, the superthermal ions can be found in both the upstream and downstream.

# I. Introduction

In a quasi-parallel shock, the angle between the upstream background magnetic field and the shock normal ($\theta_{Bn}$) is smaller than $45°$ (Jones & Ellison 1991). Due to such a peculiar property, the reflected ions by a supercritical quasi-parallel shock can travel far upstream along the magnetic field, and the resulted plasma beam instabilities can excite various large amplitude low-frequency waves (Fairfield 1969; Russell & Hoppe 1983; Gary et al. 1984; Lin 2003; Eastwood et al. 2004; Blanco-Cano et al. 2009, 2011; Wilson III et al. 2013; Omidi et al. 2013; Wu et al. 2015). These waves are then brought back by the upstream plasma towards the shock front. In such a process, the waves begin to steepen and the amplitude is enhanced when they approach the shock front, and at last a new shock front is formed after the waves interact and merge with the shock front (Burgess 1989; Thomas et al., 1990; Winske et al., 1990; Schwartz & Burgess 1991; Scholer & Burgess 1992; Scholer et al. 1993; Schwartz et al. 1992; Su et al. 2012a, 2012b). These large amplitude waves play an important role in shock diffusive acceleration (DSA) by scattering the reflected ions across a quasi-parallel shock many times, while DSA is the mechanism responsible for almost universally observed power-law spectra of energetic particles from cosmic rays to gradual solar energetic particle event (Axford et al. 1977; Bell 1978; Blandford & Ostriker 1978; Lee 1983; Zank et al. 2000; Li et al. 2003; Giacaclone 2003; Zuo et al., 2013; Cui et al., 2015). However, for a particle to be accelerated at a shock by the DSA mechanism, the particle must be sufficiently energetic to become a seed particle of DSA that it can be scattered across all the

micro- and macrostructure of the shock many times, and this is well known "injection problem" (Jokipii 1987; Zank et al. 2001; Scholer et al. 2002; Su et al. 2012a; Caprioli et al., 2015; Johlander et al., 2016).

How thermal ions in the upstream of a quasi-parallel shock become the superthermal ions and then provide the seed particles for the further acceleration by DSA have been thoroughly investigated with hybrid simulations by several authors (Scholer 1990; Scholer & Burgess 1992; Kucharek & Scholer 1991; Su et al. 2012a, 2012b; Guo & Giacalone 2013). The superthermal ions at a quasi-parallel shock come from the reflected ions by the shock after they stay close to the shock front and are accelerated for an extended period of time (Scholer 1990; Guo & Giacalone 2013). Such a process is considered as the initial state of DSA, and these ions provide the seed particles for the further acceleration by DSA (Scholer & Burgess 1992; Su et al. 2012a; Guo & Giacalone 2013). Kucharek and Scholer (1993) further found that the acceleration from the reflected ions to the superthermal ions is mainly due to grad B drift around the shock front. Su et al. (2012a; 2012b) pointed out that the extended stay close to the shock front of these reflected ions is resulted from the trapping between the new and old shock fronts during the reformation of the quasi-parallel shock. These ions are accelerated every time when they are reflected by the new shock front, and at last these ions escape to the upstream and become superthermal ions after the reformation cycle of the shock is finished. These superthermal ions can be accelerated out of the core part as well as the outer part of the velocity space of the incident upstream plasma. The other reflected ions will return to the shock immediately and then transmit to the downstream quickly, and these ions lead to ion heating in the downstream.

However, a 1-D hybrid simulation model can't take into account the influence of the structures along the shock surface on ion dynamics. Recently, the ripples with local curvature variations around the shock front are found to be inherent structures of a quasi-parallel shock, which causes fast, deflected jets in the downstream (Hietala et al. 2009; Hietala & Plaschke 2013). The characteristics of such high-speed jets in the downstream of a quasi-parallel shock has already been identified by satellite observations in the Earth's bow shock (Němeček et al. 1998; Savin et al. 2008; Archer et al. 2012; Hietala et al. 2009; Hietala et al. 2012; Plaschke et al. 2013). In this paper, by performing a two-dimensional (2-D) hybrid simulation, we investigate ion dynamics at a rippled quasi-parallel shock, and the effects of the ripples on both the reflected and transmitted ions are considered.

In this paper, we firstly give the description of the simulation model in Section II; the simulation results are presented in Section III; in Section IV, we discuss and summarize the results.

## II. Simulation Model

A 2-D hybrid simulation is performed in this paper to investigate ion dynamics at a rippled quasi-parallel shock. A hybrid simulation model treats ions as macroparticles, and their motions are governed by

$$m_p \frac{d\mathbf{v}_p}{dt} = e(\mathbf{E} + \frac{\mathbf{v}_p \times \mathbf{B}}{c}) - e\eta \mathbf{J} \qquad (1)$$

where $\mathbf{v}_p$ is the ion velocity, and $m_p$ is its mass. $\mathbf{E}$, $\mathbf{B}$ and $\mathbf{J}$ represent the electric field, magnetic field and current, respectively. $\eta$ is the resistivity resulted from the interaction between particles and high frequency waves. Electrons are treated as massless fluid, and the momentum equation is

$$-en_e(\mathbf{E} + \frac{\mathbf{V}_e \times \mathbf{B}}{c}) - \nabla p_e + en_e\eta \mathbf{J} = \frac{\partial}{\partial t}(n_e m_e \mathbf{V}_e) = 0$$

Therefore,

$$\mathbf{E} = -\frac{\mathbf{V}_e \times \mathbf{B}}{c} - \frac{\nabla p_e}{n_e e} + \eta \mathbf{J} \qquad (2)$$

where $\mathbf{V}_e$ is the bulk velocity of electrons, and $n_e$ is the number density of electrons. The electron pressure is expressed as $p_e = n_e k T_e$, where $T_e$ is the electron temperature and $k$ is the Boltzmann constant.

In the hybrid simulation model, the charge neutrality is assumed. Then, $n_e = n_p = \int f(\mathbf{v}_p) d^3\mathbf{v}_p = n$, where $n_p$ is the number density of ions and $f(\mathbf{v}_p)$ is the velocity distribution of ions, which can be obtained after we know positions and velocities of all particles. The current can be calculated with Ampere's law,

$$\frac{4\pi}{c}\mathbf{J} = \nabla \times \mathbf{B} \qquad (3)$$

Then, we can know the bulk velocity of electrons from the current ($\mathbf{J}$) and the bulk velocity of ions ($\mathbf{V}_p$, where $\mathbf{V}_p = \int \mathbf{v}_p f(\mathbf{v}_p) d^3\mathbf{v}_p$) according to the equation $\mathbf{J} = ne(\mathbf{V}_p - \mathbf{V}_e)$.

The magnetic field can be calculated from Faraday's law:

$$\frac{1}{c}\frac{\partial \mathbf{B}}{\partial t} = -\nabla \times \mathbf{E} \tag{4}$$

and

$$\nabla \cdot \mathbf{B} = 0 \tag{5}$$

Besides Eqs. (1)-(5), we still need the equation of state of the electrons, which is assumed to be adiabatic, to implement the algorithm of the hybrid simulation model, and the details can be found in Winske (1985).

Initially, plasma with a fixed bulk velocity ($V_{inj} = 4.5 V_A$, $V_A$ is the upstream Alfven speed) moves to right rigid boundary, and the background magnetic field $B_0$ lies in the $x-y$ plane. The plasma is reflected when it gets the right boundary and interacts with continuous injected plasma, and this interaction leads to the formation of shock front. Meanwhile, the shock front has a propagating velocity pointing to the left along the $x$ direction, which is the global shock normal. A periodic boundary condition is used in the $y$ direction. For the shock in this simulation, $\theta_{Bn}$ ($\theta_{Bn}$ is the angle between the shock normal and direction of upstream background magnetic field) is $30°$, and the upstream plasma beta is $\beta_p = \beta_e = 0.4$ (where $p$ and $e$ indicate proton and electron, respectively). The velocity of shock in the downstream reference frame is about $1.0V_A$, and then the Mach number is around 5.5, which is a typical value for a terrestrial bow shock. The simulation plane cover an area with length $L_x = n_x \times \Delta x = 1000 \times 0.5 c/\omega_{pi} = 500 c/\omega_{pi}$ and width $L_y = n_y \times \Delta y = 300 \times 1.0 c/\omega_{pi} = 300 c/\omega_{pi}$ (where $n_x$ and $n_y$ are the number of grid cell, $\Delta x$ and $\Delta y$ are grid sizes, $c$ is the speed of light and $\omega_{pi}$ is the ion plasma frequency under upstream parameters). The electron resistive length ($L_\eta = \eta c^2/(4\pi V_A)$, where $\eta$ denotes the wave-particle effects resulting from high frequency plasma instabilities and $c$ is the light speed) is set to be $L_\eta = 0.1$, which is much smaller than the grid size. The time step is $\Delta t = 0.02 \Omega_i^{-1}$ (where $\Omega_i = eB_0/m$ is the ion gyro-frequency).

## III. Simulation Results

Ripples with the local curvature variations around the shock front are inherent structures of quasi-parallel shock, and they can be seen clearly in Figure 1, which plots the total magnetic field at $\Omega_i t = 125$. The shock front is around $x = 364 c/\omega_{pi}$, where obvious ripples with the size about 75 $c/\omega_{pi}$ can be found along the $y$ direction. Here the position of shock is calculated as follows: we average the magnetic field along the $y$ direction and obtain a one-dimensional shock, and at such a one-dimensional shock the total magnetic field have the maximum gradient at the position of shock. The plasma waves with the amplitude $\delta B/B_0 \sim 1$ and wavelength about 50 $c/\omega_{pi}$ exist in the upstream. These waves have also been identified in previous simulations (Scholer & Burgess 1992; Scholer et al. 1993; Scholer 1993; Blanco-Cano et al. 2009; Su 2012a, 2012b), and they correspond to the reported ultra-low frequency (ULF) waves in satellite observations (Schwartz et al. 1992; Burgess et al. 2005; Lucek et al. 2008; Eastwood et al. 2005a, 2005b; Wu et al. 2015). The amplitude of these waves will become large as they approach the shock front, and at last a new shock front may be generated. Such a reformation of the shock can be demonstrated more clearly in Figure 2, which shows the time evolution of the total magnetic field along $y=(a)118c/\omega_{pi}$, (b)$153c/\omega_{pi}$, (c)$195c/\omega_{pi}$ and (d) $228c/\omega_{pi}$, as denoted by the dashed lines in Figure 1. The period of the shock reformation is about $25\Omega_i^{-1}$. Due to the existence of the ripples around the shock front, the reformation is not synchronous at different $y$ positions.

By following a group of ions, we can investigate ion dynamics at the rippled quasi-parallel shock. These particles are located in the area ($320c/\omega_{pi} \leq x \leq 370c/\omega_{pi}$, $114c/\omega_{pi} \leq y \leq 141c/\omega_{pi}$) and ($320c/\omega_{pi} \leq x \leq 380c/\omega_{pi}$, $168c/\omega_{pi} \leq y \leq 198c/\omega_{pi}$) at $\Omega_i t=121.5$, which are denoted by "A" and "B" in Figure 3. At $\Omega_i t=121.5$, these particles just begin to interact with the shock front. The particles in the region "A" interact with the lower part of one selected ripple, while the particles in the region "B" interact the upper part of the same ripple. Figure 4 and 5 respectively plot the evolution of the ions in the regions "A" and "B" (These particles at $\Omega_i t=121.5$ are restricted in the regions "A" and "B" denoted by the red boxes in Figure 3) at four different times $\Omega_i t=121.5$, 161.5, 171.5 and $\Omega_i t=181.5$. In both the figures, the left columns represent the positions of particles (the magnetic field is also plotted for reference), the middle columns describe the corresponding velocity distribution ($v = \sqrt{v_x^2 + v_y^2 + v_z^2}$), and the right columns present the distribution of the ion velocity in the $x$ direction ($v_x$). In the left columns, the red particles mean that these particles are located in the upstream of shock, and the position of shock front is indicated by the red dashed lines, which are located at $x = 338c/\omega_{pi}$, $348c/\omega_{pi}$ and $358c/\omega_{pi}$ at $\Omega_i t = 161.5$, $\Omega_i t = 171.5$ and $\Omega_i t = 181.5$, respectively (the position of shock is calculated with the same method described in Figure 1). In the middle and right columns, the red area means the percentage of the particles located in the upstream of shock. At $\Omega_i t = 121.5$, the particles begin to interact with the shock, and the interaction time lasts for about $4.5\,\Omega_i^{-1}$ (The interaction time is the time period from the time when the particles just begin to interact with the shock to the time when the interaction almost finishes), which is much smaller than the period of shock reformation ($\sim 25\Omega_i^{-1}$). Therefore, during the interaction between the particles and shock, the shock can be considered as stationary. After the interaction, the particles can be separated into two parts: the particles transmitted into the downstream and the others remains in the upstream, and both of these particles almost move along the magnetic field. Obviously, the ions in the region "A" (in the lower part of the ripple) are easier to be reflected by the shock. As shown in right columns, the particles remains in the upstream can have either negative $v_x$ or positive $v_x$, which means that these particles can move toward the upstream or downstream of the shock. The reason that the particles remaining in the upstream can move toward the downstream is due to the scattering of the upstream waves, as discussed in the follows by tracing the trajectories of several typical ions. At last, for the particles from the region "A", the particles remains in the upstream occupy about 1.48% of the total particles. For the particles from the region "B", the percentage of the particles remaining in the upstream is about 0.22%. In region "A", the local shock can be considered as a quasi-parallel shock because the local shock angle (the angle between the local shock normal and upstream magnetic field) is less than $45^0$, and the upstream particles are reflected by the shock more efficiently and easy to escape to the upstream. In region "B", the local shock is more like a quasi-perpendicular shock, and the upstream particles are more difficult to be reflected and escape to the upstream. From the velocity distribution of particles, we can find that the ions are highly accelerated with the maximum velocity $\sim 30V_A$ after they interact with the shock. Although most of the superthermal ions comes from the ions remaining in the upstream after they interact with the shock, we can still find that a part of superthermal ions come from the downstream. This is different from the 1-D hybrid simulation results of a quasi-parallel shock (Su et al. 2012a), where all of superthermal ions come from the upstream, and this will be demonstrated in the follows by tracing the trajectories of several typical ions. There is also no surprise that there are more superthermal ions from the region "A" than those from the region "B", because the local shock in the region "A" is more like a quasi-parallel shock. For the particles which transmit into the downstream, their bulk velocity of the particles from the region "B" is larger than that from the region "A". Therefore, the lower part of the ripple tends to not only reflect more upstream ions, but also decelerate more efficiently the transmitted ions.

Because of the different characteristics of transmitted ions after the upstream ions interact with the different parts of a ripple at a quasi-parallel shock, the bulk velocities of these transmitted ions are different along the $y$ direction, which forms high-speed jets observed in the downstream of a quasi-parallel shock by Cluster (Hietala et al. 2009). In order to demonstrate the generation mechanism of the high-speed jets, in Figure 6 we plot (a) the total magnetic field $B/B_0$, (b) the bulk velocity $V/V_A$, (c) the local angle between the magnetic field and the $x$ direction $\theta_{Bx}$, and (d) the electric field in the $x$ direction $E_x$ at $\Omega_i t = 121.5$, when the particles from the regions "A" and "B" just begin to interact with the shock. Here "PA" and "PB" denote the lower and upper parts of the ripple. A high-speed jet in the just downstream corresponding to the upper part of the ripple, which is denoted by the yellow arrow in Figure 6(b), can be easily identified. The bulk velocity of the particles in a high speed jet is larger than that in the other downstream areas around it, and its size is about $15\,c/\omega_{pi}$. Comparing with the lower part of the ripple (denoted by "PA"), $\theta_{Bx}$ is smaller and the electric field $E_x$ is positive in the upper part of the ripple (denoted by "PB"), where the particles are decelerated with a less efficiency when they transmit through the shock and lead to the formation of high-speed jets in the corresponding downstream.

In 1-D hybrid simulations of a quasi-parallel shock, all of superthermal ions come from the upstream, where these particles are reflected by the shock and get accelerated when they are trapped between the old and new shock fronts until escape to the upstream (Su et al., 2012a). Here, in 2-D hybrid simulations of a quasi-parallel shock, superthermal ions come from not only the upstream but also the downstream, as shown in Figure 4. In order to investigate dynamics of these superthermal ions, we follow the trajectories of these particles, and find that there are four different categories. Figure 7 plots a typical ion trajectory of the first category. Figure 7(a), (b) and (c) show the typical trajectory in the ($x$, $y$) plane at $\Omega_i t = 108.0 \sim 143.0$, $143.0 \sim 151.5$ and $151.5 \sim 188.5$, while 7(d) presents the evolution of its kinetic energy. The total magnetic fields at $\Omega_i t = 143.0$, $\Omega_i t = 151.5$ and $\Omega_i t = 180.0$ are overlaid in Figure 7(a), (b) and (c) for reference. At $\Omega_i t = 108.0$, the particle is in the upstream, and moves toward the shock. It is reflected by the shock at "A2", and then moves along the surface of shock. At "A3", it leaves the shock and goes to the upstream, however, it is trapped by a wave from the upstream. The wave is steepening when approaching the shock front, and becomes a new shock front until it merges with the old shock front. At "A5", the particle again leaves the shock, and crosses the upstream waves until goes to the further upstream. Please note, here the particle crosses the lower boundary will enter the simulation domain from the upper boundary due to the periodic boundary condition used in the simulation. The particle suffers two acceleration stages: In the first stage (from "A2" to "A3", the particle is reflected by the shock when approaching to the shock front from the upstream and moves along the shock surface; In the second stage (from "A3" to "A5"), the particle is trapped between new and old shock fronts. The acceleration process is similar to that in 1-D quasi-parallel shock, which has been demonstrated clearly by Su et al. (2012a), except that now in 2-D simulations the particle can move along the shock surface and escape easily to the upstream due to the inhomogeneity of the shock front and upstream waves along the $y$ direction.

Figure 8 plots a typical ion trajectory of the second category. Figure 8(a), (b) and (c) show the typical trajectory in the ($x$, $y$) plane at $\Omega_i t = 110.5 \sim 140.5$, $140.5 \sim 149.5$ and $149.5 \sim 174.5$, while 8(d) presents the evolution of its kinetic energy. The total magnetic fields at $\Omega_i t = 140.5$, $\Omega_i t = 149.5$ and $\Omega_i t = 174.5$ are overlaid in Figure 8(a), (b) and (c) for reference. The particle from the upstream is reflected by the shock, and then it is accelerated when moving along the shock surface and trapped between the new and old shock fronts. However, different from the particles from the first category, the particle at last enters the downstream after crossing the shock front where the magnetic field is weak

Figure 9 plots a typical ion trajectory of the third category. Figure 9(a), (b), (c) and (d) show the typical trajectory in the ($x$, $y$) plane at $\Omega_i t = 114.0 \sim 144.0$, $144.0 \sim 158.0$, $158.0 \sim 172.0$, and $172.0 \sim 188.5$, while 9(e) presents the evolution of its kinetic energy. The total magnetic fields at $\Omega_i t = 144.0$, $\Omega_i t = 158.0$, $\Omega_i t = 172.0$, and $\Omega_i t = 188.5$ are overlaid in Figure 8(a), (b),(c) and (d) for reference. After reflected by the shock, the particle is accelerated when moving along the shock surface and trapped between the new and old shock fronts. However, after the particle escape to the upstream, it can be trapped again by the wave in the further upstream and suffers acceleration process more than two stages. At last, the particle may escape to the further upstream (see Figure 9), or cross the shock and go to the downstream (see Figure 10, which plots a typical ion trajectory of the fourth category). Figure 10(a), (b), (c) and (d) show the typical trajectory in the ($x$, $y$) plane at $\Omega_i t = 112.5 \sim 142.5$, $142.5 \sim 151.5$, $151.5 \sim 165.5$, and $165.0 \sim 188.5$, while 10(e) presents the evolution of its kinetic energy. The total magnetic fields at $\Omega_i t = 142.5$, $\Omega_i t = 151.5$, $\Omega_i t = 165.5$, and $\Omega_i t = 188.5$ are overlaid in Figure 10(a), (b),(c) and (d) for reference. The particle at last go to downstream after it suffers acceleration process more than two stages.

In summary, the superthermal ions are those particles reflected by the shock and get accelerated when moving along the shock surface and then trapped between the upstream waves (or new shock front) and the shock front. At last, those particles may escape to the further upstream or enter the downstream. Therefore, the superthermal ions can be found in both the upstream and downstream. About 20% of the superthermal ions are found in the downstream, and 30% suffer acceleration process more than two stages. In general, the particles can be accelerated to much higher energy if they suffers acceleration process more than two stages than those only undergoing two-stage acceleration process, and it means that the trapping in upstream waves and subsequent acceleration are more efficient than trapping in the new shock front alone. The importance of the upstream waves in the generation of seed particles for further DSA acceleration has already been emphasized with satellite observations (Wu et al. 2015, Johlander et al. 2016).

# IV. Conclusions and Discussion

In this paper, 2-D hybrid simulations are performed to investigate ion dynamics at a quasi-parallel shock. Obvious ripples are found to form around the shock front, at the same time, the shock is reforming: a new shock front may appear in the upstream, which is convected to the shock front by the upstream flow, and at last merges with the old shock front. However, the reformation of the shock is not synchronous along the $y$ direction due to the existence of the ripples. When the upstream ions interact with the quasi-parallel shock, their behaviors will be different at different $y$ direction. The upstream ions tend to be reflected in the lower part of a ripple, while they are easier to transmit through the upper part of a ripple. In the downstream corresponding to the upper part of a ripple, the bulk velocity is larger, and then a high-speed jet is formed. Therefore, the observed high-speed jets downstream are the results of the ripples inherent in a quasi-parallel shock. High-speed jets have already been observed by satellites in the downstream of quasi-parallel shocks (Němeček et al. 1998; Archer et al. 2012; Hietala et al. 2012; Plaschke et al. 2013). The upstream particles tend to be reflected by the shock in the lower part of a ripple. The reflected ions by the shock are accelerated when they move along the shock surface, or are trapped between the upstream waves (include the new shock front) and the shock front.

The particles accelerated by a shock through DSA mechanism is considered to be one of the important sources for observed power-law spectra of energetic particles from cosmic rays to gradual solar energetic particle event. However, in order that DSA mechanism works in a shock, the energy of the particles must exceed a threshold, or the particles need at first to be pre-accelerated to become superthermal particles. The acceleration of the reflected particles by a shock is considered to provide such a pre-acceleration mechanism. In 1-D simulations of a quasi-parallel shock, the reflected ions are accelerated when they are trapped between the new and old shock fronts during the reformation of the shock. These particles escape from the shock to the upstream after the cycle of reformation is finished, and the superthermal ions can only be found in the upstream (Su et al. 2012a). Here, in 2-D simulations of a quasi-parallel shock, because of the inhomogeneity of shock front and upstream waves along the $y$ direction, the reflected ions may be leaked to the upstream or the downstream when they are trapped between the old and new shock fronts. The leaked particles to the upstream may be accelerated again due to the interaction with the upstream waves. The superthermal ions can be found in both the upstream and downstream, and they may suffer multiple stage several acceleration.

## Acknowledgments

This work was supported by the National Science Foundation of China, Grant Nos. 41331067, 11235009, 41527804, 41474125, 41421063, 973 Program (2012CB825602, 2013CBA01503).

**Figure Captions**

Figure 1. The total magnetic field at $\Omega_i t = 125$. The dashed lines denote four cuts along y = 118, 153, 195 and $228 c/\omega_{pi}$. The red arrow indicates the direction of the upstream magnetic field.

Figure 2. The evolution of the total magnetic field along (a) $y = 118 c/\omega_{pi}$, (b) $y = 153 c/\omega_{pi}$, (c) $y = 195 c/\omega_{pi}$, (d) $y = 228 c/\omega_{pi}$, which have been denoted by dashed lines in Figure 1.

Figure 3. The position of the selected ions at $\Omega_i t = 121.5$, whose positions are restricted in the area ($250 \leq x \leq 270$, $0 \leq y \leq 300$) at $\Omega_i t = 100$. The boxes "A" and "B" denote the areas ($310 c/\omega_{pi} \leq x \leq 370 c/\omega_{pi}$, $114 c/\omega_{pi} \leq y \leq 141 c/\omega_{pi}$) and ($315 c/\omega_{pi} \leq x \leq 380 c/\omega_{pi}$, $168 c/\omega_{pi} \leq y \leq 198 c/\omega_{pi}$), respectively.

Figure 4. The time evolution of the particles in areas "A" at $\Omega_i t = 121.5$, 161.5, 171.5 and 181.5. The particles at $\Omega_i t = 121.5$ are restricted in areas "A", which have been indicated in Figure 3. The left column represents the positions of particles (The magnetic field are plotted for reference), and the red particles mean that these particles are located in the upstream of the shock. The position of shock front is indicated by the red dashed line in the left column, which are $338 c/\omega_{pi}$, $348 c/\omega_{pi}$ and $358 c/\omega_{pi}$ at $\Omega_i t = 161.5$, $\Omega_i t = 171.5$ and $\Omega_i t = 181.5$, respectively. The middle column plots the corresponding velocity distribution ($v = \sqrt{v_x^2 + v_y^2 + v_z^2}$), (where $N_t$ is the total number of particles, and $N$ is the number in a definite velocity range), and the right column shows the distribution of the ion velocity in the $x$ direction ($v_x$). The red area means the percentage of the particles located in the upstream of the shock.

Figure 5. The time evolution of the particles in areas "B" at $\Omega_i t = 121.5$, 161.5, 171.5 and 181.5. The particles at $\Omega_i t = 121.5$ are restricted in areas "B", which have been indicated in Figure 3. The left column represents the positions of particles (The magnetic field are plotted for reference), and the red particles mean that these particles are located in the upstream of the shock. The position of shock front is indicated by the red dashed line in the left column, which are $338 c/\omega_{pi}$, $348 c/\omega_{pi}$ and $358 c/\omega_{pi}$ at $\Omega_i t = 161.5$, $\Omega_i t = 171.5$ and $\Omega_i t = 181.5$, respectively. The middle column plots the corresponding velocity distribution ($v = \sqrt{v_x^2 + v_y^2 + v_z^2}$), (where $N_t$ is the total number of particles, and $N$ is the number in a definite velocity range), and the right column shows the distribution of the ion velocity in the $x$ direction ($v_x$). The red area means the percentage of the particles located in the upstream of the shock.

Figure 6. (a) the total magnetic field $B/B_0$, (b) the bulk velocity $V/V_A$, (c) the local angle between the magnetic field and the $x$ direction $\theta_{Bx}$, and (d) the electric field in the $x$ direction $E_x$ at $\Omega_i t = 121.5$, when the particles from the regions "A" and "B" just begin to interact with the shock. Here "PA" and "PB" denote the lower and upper parts of a ripple. A yellow bold arrow in right upper panel labels a high-speed jet in the downstream of the shock.

Figure 7. A typical ion trajectory of the first category. (a), (b) and (c) show the typical

trajectory in the ($x$, $y$) plane at $\Omega_i t = 108.0\sim143.0$, $143.0\sim151.5$ and $151.5\sim188.5$, while (d) presents the evolution of its kinetic energy. The total magnetic fields at $\Omega_i t = 143.0$, $\Omega_i t = 151.5$ and $\Omega_i t = 180.0$ are overlaid in (a), (b) and (c) for reference. The two areas denoted "I" ands "II" show two accelerating stages.

Figure 8. A typical ion trajectory of the second category. (a), (b) and (c) show the typical trajectory in the ($x$, $y$) plane at $\Omega_i t = 110.5\sim140.5$, $140.5\sim149.5$ and $149.5\sim174.5$, while (d) presents the evolution of its kinetic energy. The total magnetic fields at $\Omega_i t = 140.5$, $\Omega_i t = 149.5$ and $\Omega_i t = 174.5$ are overlaid in (a), (b) and (c) for reference. The two areas denoted "I" ands "II" show two accelerating stages.

Figure 9. A typical ion trajectory of the third category. (a), (b), (c) and (d) show the typical trajectory in the ($x$, $y$) plane at $\Omega_i t = 114.0\sim144.0$, $145.0\sim158.0$, $158.0\sim172.0$, and $172.0\sim188.5$, while (d) presents the evolution of its kinetic energy. The total magnetic fields at $\Omega_i t = 144.0$, $\Omega_i t = 158.0$, $\Omega_i t = 172.0$, and $\Omega_i t = 188.5$ are overlaid in (a), (b),(c) and (d) for reference. The two areas denoted "I" , "II" ands "II" show three accelerating stages.

Figure 10. A typical ion trajectory of the fourth category. (a), (b), (c) and (d) show the typical trajectory in the ($x$, $y$) plane at $\Omega_i t = 112.5\sim142.5$, $142.5\sim151.5$, $151.5\sim165.5$, and $165.5\sim188.5$, while (d) presents the evolution of its kinetic energy. The total magnetic fields at $\Omega_i t = 142.5$, $\Omega_i t = 151.5$, $\Omega_i t = 165.5$, and $\Omega_i t = 188.5$ are overlaid in (a), (b),(c) and (d) for reference. The two areas denoted "I" , "II" ands "II" show three accelerating stages.

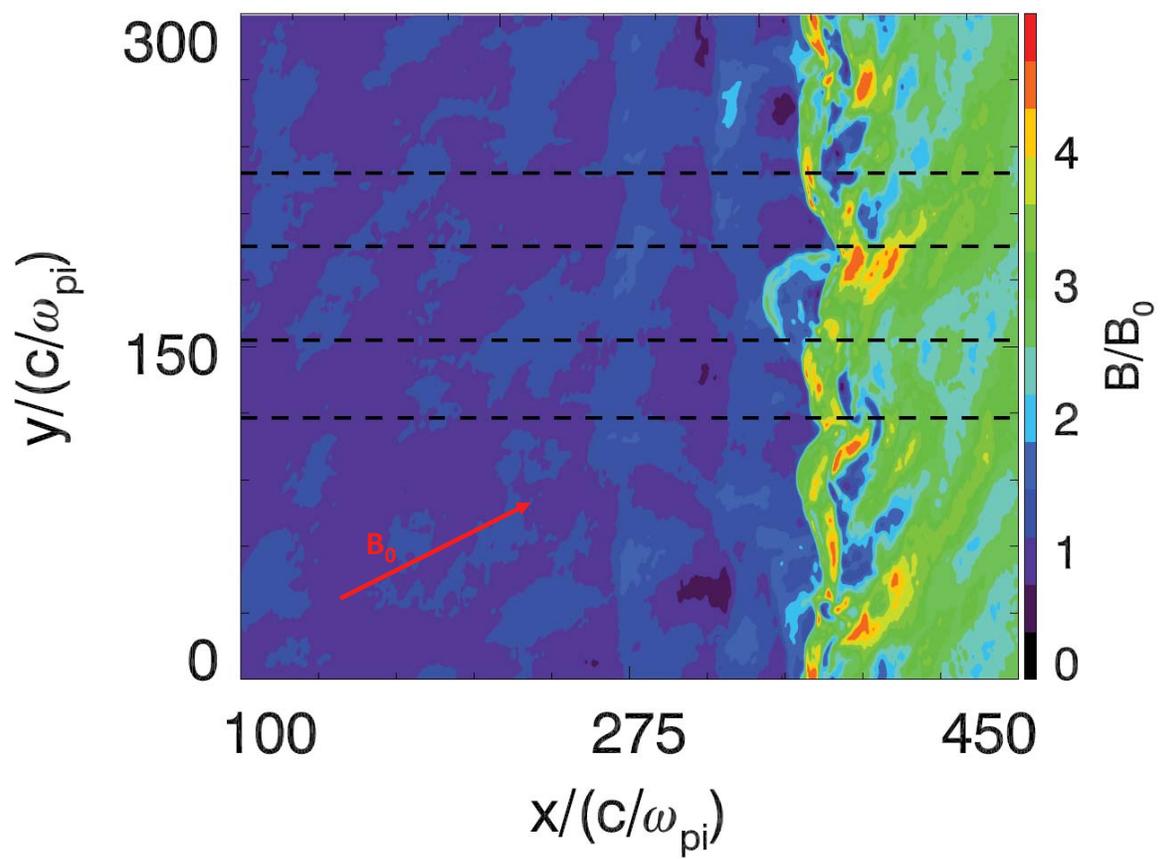

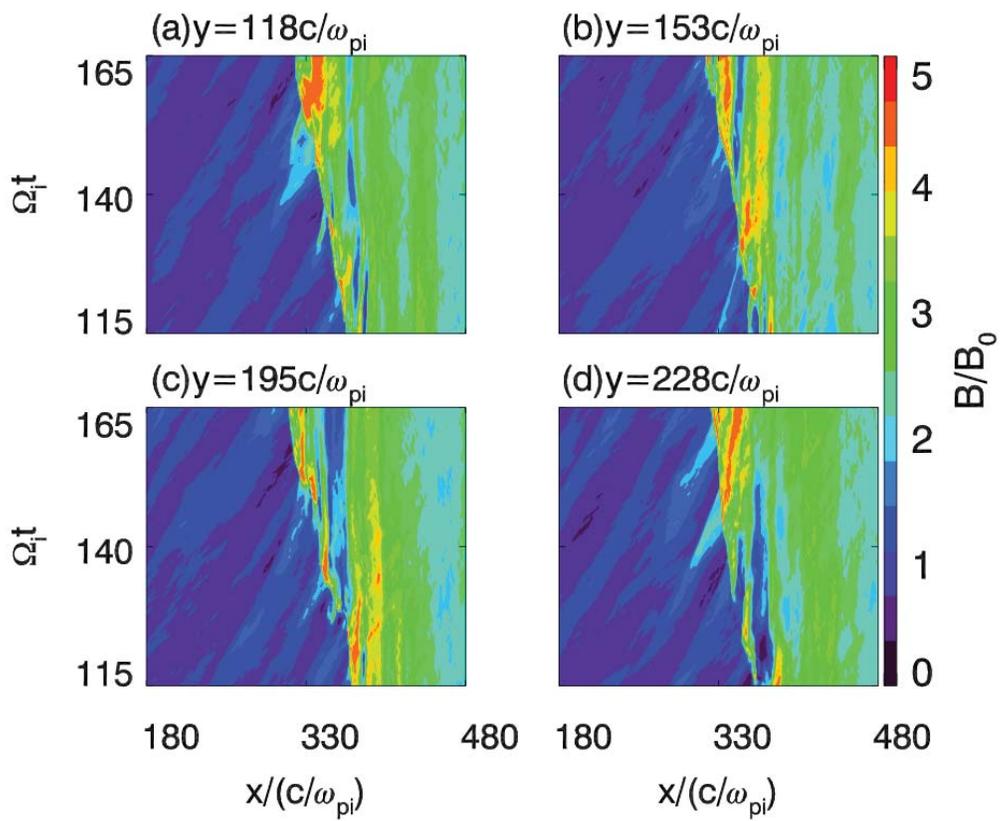

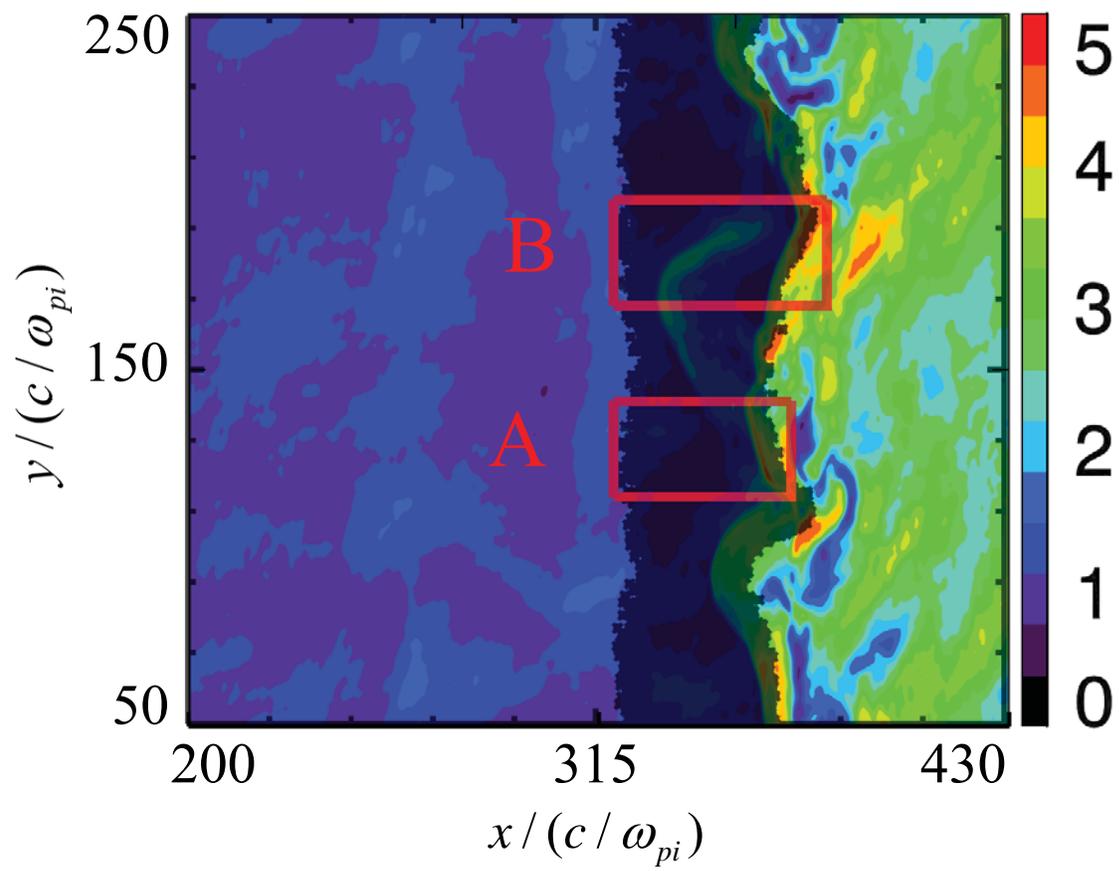

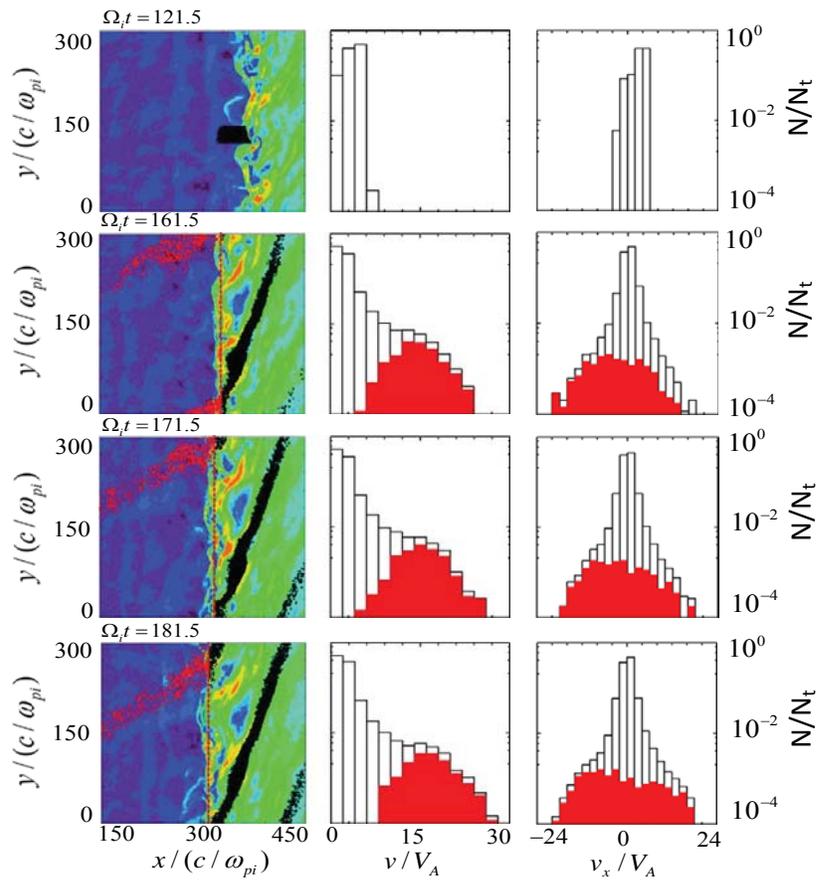

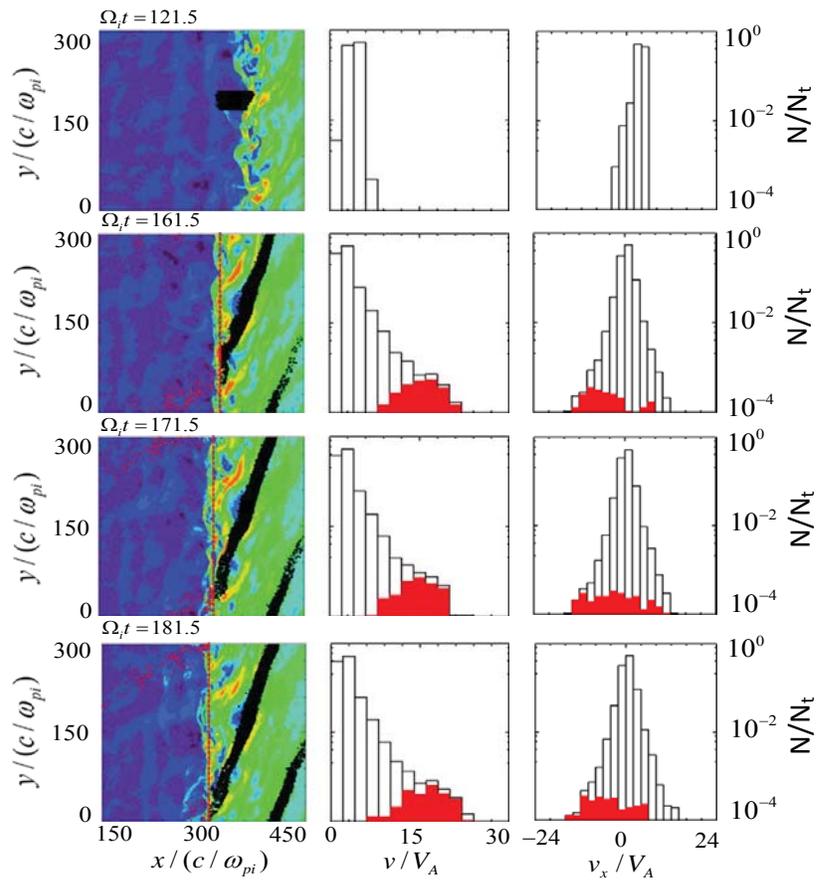

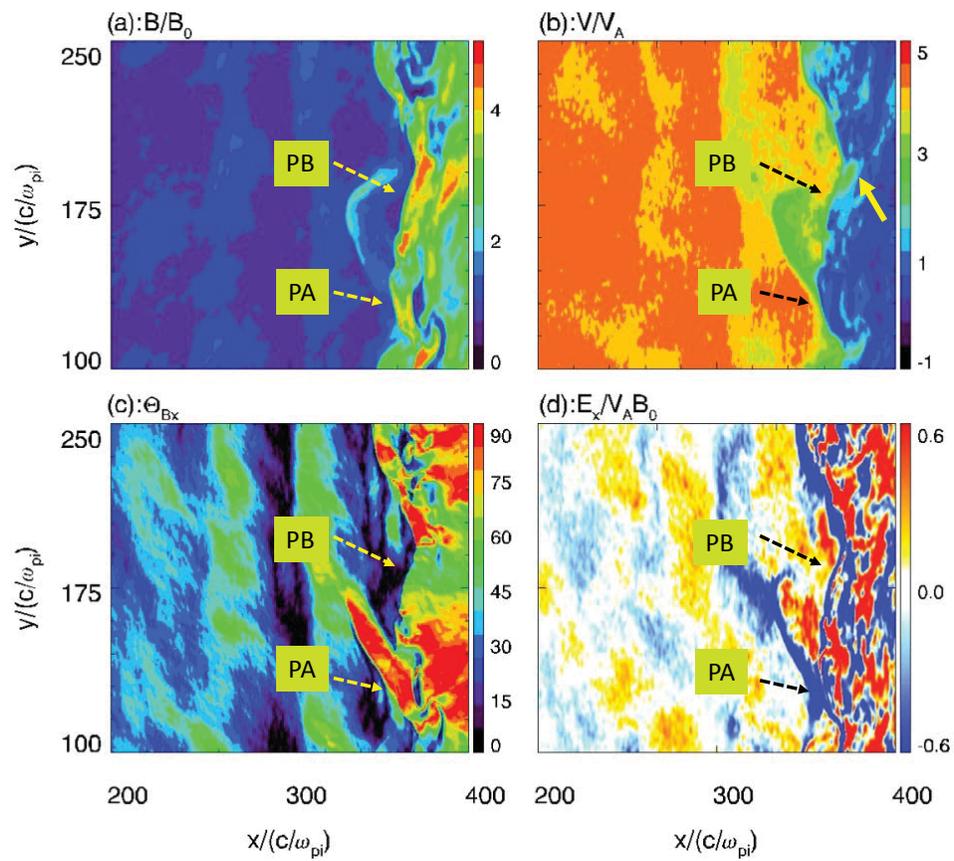

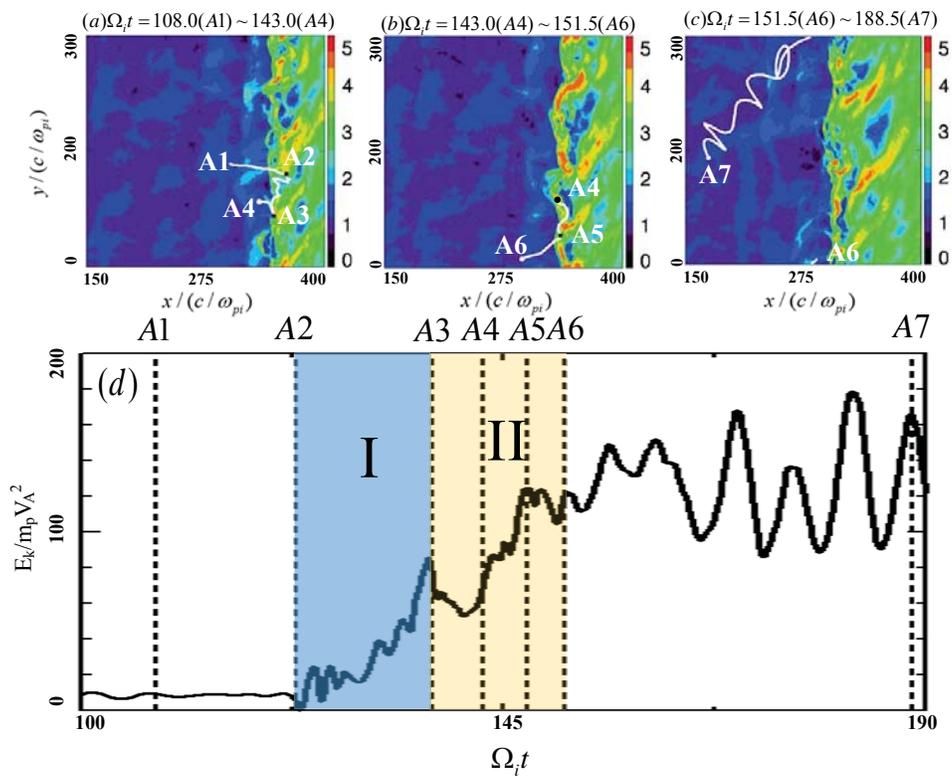

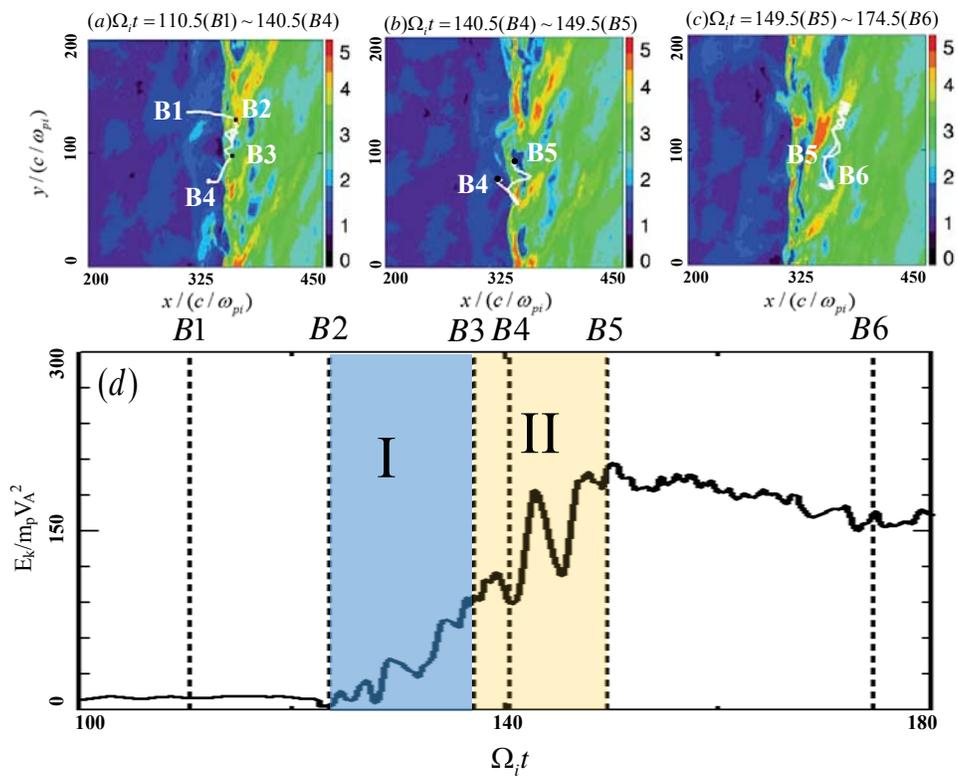

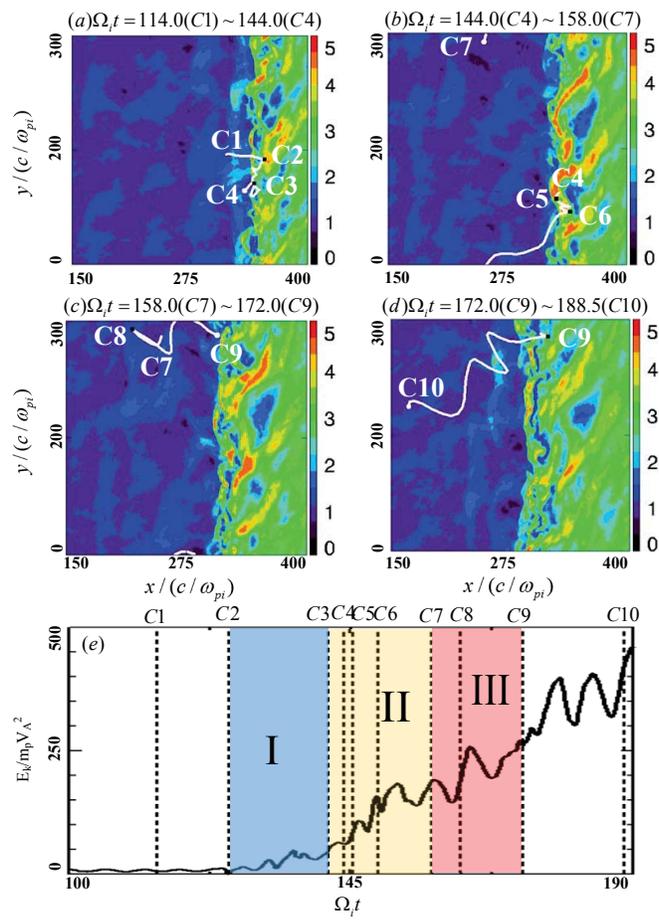

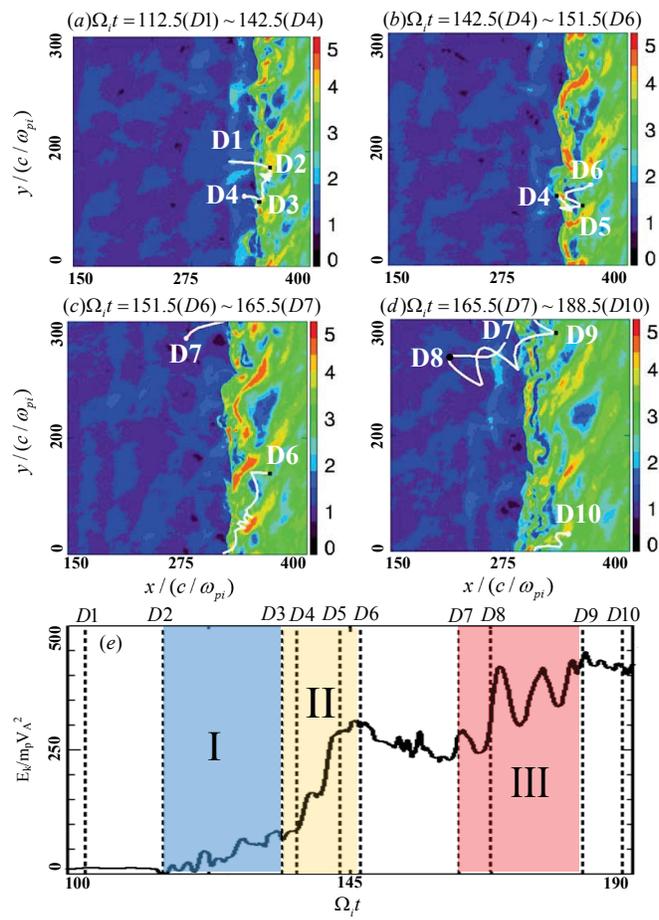